\documentclass[twocolumn,showpacs,preprintnumbers,amsmath,amssymb]{revtex4}
\usepackage{graphicx}% Include figure files
\usepackage{dcolumn}% Align table columns on decimal point
\usepackage{bm}% bold math

\begin{document}

\title{Complex Effective Action and Schwinger Effect}

\author{Sang Pyo Kim}
\affiliation{Department of Physics, Kunsan National University, Kunsan 54150, Korea}
\email{sangkim@kunsan.ac.kr}

\begin{abstract}
Spontaneous pair production from background fields or spacetimes is one of the most prominent phenomena predicted by quantum field theory. The Schwinger mechanism of production of charged pairs by a strong electric field and the Hawking radiation of all species of particles from a black hole are the consequence of nonperturbative quantum effects. In this review article, the vacuum structure and pair production is reviewed in the in-out formalism, which provides a consistent framework for quantum field theory in the sense that the complex action explains not only the vacuum persistence but also pair production. The current technology of intense lasers is still lower by a few order than the Schwinger limit for electron-positron pair production, while magnetic fields of magnetars on the surface are higher than the Schwinger limit and even higher at the core. On the other hand, the zero effective mass of electron and hole in graphene and Dirac or Weyl semimetals will open a window for experimental test of quantum electrodynamics (QED) phenomena in strong fields.
\end{abstract}
\date{\today}
%\pacs{98.80.Cq, 04.62.+v, 12.20.-m, 03.65.Pm}

\maketitle

\section{Introduction}\label{sec1}

The Schwinger mechanism of production of charged pairs due to an electric field \cite{schwinger51}, the Hawking radiation of all species of particles from a black hole \cite{hawking75} and the cosmic particle production due to an expanding spacetime \cite{parker68} are the most prominent nonperturbative quantum effect in quantum field theory. Soon after the discovery of the Dirac theory, Sauter found that the Dirac vacuum became unstable against the pair creation of electrons and positrons \cite{sauter31}. Heisenberg and Euler then found the one-loop effective action for an electron in a constant electromagnetic field, which is now known as the Heisenberg-Euler or Schwinger effective action \cite{heisenberg-euler36}. It is remarkable that the interaction of an electron and an electromagnetic field background has been studied nonperturbatively even before the advent of quantum electrodynamics. It was Schwinger who introduced the proper-time integral to express the one-loop effective action in scalar and spinor QED \cite{schwinger51}.

The most distinct feature of the Heisenberg-Euler and Schwinger action is the existence of poles in the proper-time representation of the action in the presence of an electric field. This implies that the one-loop effective action has not only the vacuum polarization (the real part) but also the vacuum persistence (twice the imaginary part). The vacuum persistence is the consequence of the spontaneous production of charged pairs from the Dirac vacuum. In other words, the Dirac sea becomes unstable against quantum tunneling of virtual particles through a tilted barrier due to an electric field and creates pairs as particle-holes.

In modern quantum field theory, particle or pair production is a consequence of different vacua in the in- and the out-regions due to a background such as an electromagnetic field, a spacetime with a horizon, and a nonstationary gravitational interaction. A uniform electric field tilts the Dirac sea due to the electrostatic potential and an electron in the Dirac sea quantum tunnels the mass barrier to become a pair of real electron and positron or a positive frequency solution in the remote past is scattered by a nonstationary potential into not only a positive frequency solution but also the negative frequency solution in the remote future, of which the amplitude of negative frequency solution measures the mean number of produced pairs a la Bogoliubov transformation between the in- and the out-vacua \cite{birrell-davies}.

In this review article we shall focus on the Schwinger mechanism and the vacuum polarization in a constant electric field with or without a constant magnetic field. Because of the conservation of energy-momentum and charge, the electrostatic potential energy of an electron over one Compton wavelength equaling to the rest mass gives the Schwinger limit, $E_c = m^2/e = 1.3 \times 10^{16} {\rm V/cm}$. The current technology of extremely intense lasers is still a few order lower than the Schwinger limit. In the near future, Extreme Light Infrastructure (ELI) will provide a few of ten PW laser beams, which still belongs to a sub-Schwinger limit \cite{DMHK12}. In the future, International Center for Zetta-Exawatt Science and Technology (IZEST) will reach or go beyond the Schwinger limit, which will spontaneously create electron-electron pairs and even fundamental particle-antiparticle pairs. On the other hand, ultramagnetized neutron stars, the so-called magnetars, have magnetic fields on the surface two or three order stronger than the Schwinger limit, $B_c = m^2/e = 4.4 \times 10^{13} {\rm G}$ \cite{duncan-thompson} and even much higher at the core. When an electromagnetic field is far greater than the Schwinger limit, that is, supercritical, the QED phenomena become effectively massless.

Graphene, a two-dimensional hexagonal array of carbons, is described by an effective theory for massless electron-hole pairs. Graphene has two equivalent $K^{\pm}$ Dirac points, whose nonequivalence produces the left- and the right-moving pseudo-spins, behaving like spin-1/2 fermions. At the Dirac point, the particle has the zero mass and ultra-relativistic dispersion relation $\epsilon = v_{\rm F} \vert k \vert$ with the Fermi velocity $v_{\rm F}$. Then, the Hamiltonian is described by the Weyl spinor. The recently discovered Dirac semimetals \cite{young12,wang12} and Weyl semimetals \cite{burkov11} are three-dimensional analogy of two-dimensional graphene. Including the time dimension, the effective theory for Dirac semimetals is prescribed by massless QED in the four-dimensional spacetime \cite{kim-saemulli}. In this sense, the massless QED may be tested by the condensed matter analogy. The characteristic feature of QED is compared with the condensed matter analogy in Table \ref{tab1}.

\begin{table}
\caption{Comparison of QED/QCD and condensed matter}\label{tab1}
\begin{ruledtabular}
\begin{tabular}{llll}
QED/QCD & condensed matter\\ \hline
Dirac or Yang-Mills theory & Dirac or Weyl theory \\
external gauge fields & external electromagnetic fields \\
Schwinger effect & Landau-Zener tunneling\\
\end{tabular}
\end{ruledtabular}
\end{table}

In this paper, we review the effective QED action and the Schwinger effect in a constant electric and magnetic field. To compute the QED action and to explore the relation between the QED action and the Schwinger effect, we shall employ the in-out formalism, which has been recently elaborated since the seminal works by Schwinger and DeWitt \cite{schwinger51b,schwinger51c,dewitt03}. We also introduce a method, such as the reconstruction conjecture, which enables one to find the effective action from the knowledge of the vacuum persistence. There are different approaches to the vacuum polarization and the Schwinger effect as summarized in Fig. 1.

\section{S-Matrix for QED Action} \label{sec2}

A spin-1/2 fermion interacting with an external electromagnetic field is prescribed by the Lagrangian density for the Dirac equation
\begin{eqnarray}
{\cal L} = \frac{i}{2} \bigl[ \bar{\psi} \gamma_{\alpha} D_{\alpha} \psi - (D_{\alpha} \bar{\psi}) \gamma_{\alpha} \psi \bigr] - m \bar{\psi} \psi .
\end{eqnarray}
where $D_{\alpha} = \partial_{\alpha} - i q A_{\alpha}$. The conventional method for computing the QED action is the $S$-matrix or the evolution operator $\hat{U}$, which is given by
\begin{eqnarray}
\hat{S} &=& {\rm T} \exp \Bigl[- iq \int d^4x \bar{\psi} \gamma_{\alpha} \psi A_{\alpha}  \Bigr] \nonumber\\
&=& {\rm T} \exp \Bigl[i \int d^4 x {\cal L}_I \Bigr].
\end{eqnarray}
Then, the out-state is expressed as
\begin{eqnarray}
\vert {\rm out} \rangle = \hat{S}^{\dagger} \vert {\rm in} \rangle, \quad \vert t \rangle = \hat{U}(t, - \infty) \vert {\rm in} \rangle,
\end{eqnarray}
and the vacuum persistence amplitude is
\begin{eqnarray}
S_0 (A) = \langle 0, {\rm out} \vert 0, {\rm in} \rangle =  \langle 0, {\rm in} \vert \hat{S} \vert 0, {\rm in} \rangle.
\end{eqnarray}

\begin{figure}\label{fig1}
\flushleft
\includegraphics[width=4.2in]{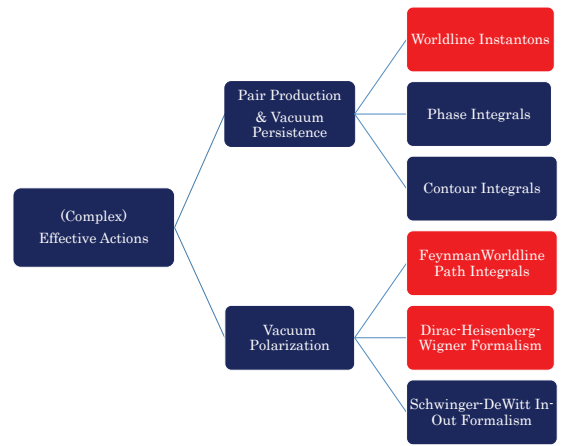}
\caption{The summary of different methods to the vacuum polarization and the Schwinger effect. It does not exhaust all known methods in literature.}
\end{figure}

By performing the path integral, the vacuum persistence amplitude for a general electromagnetic field is given by
\begin{eqnarray}
S_0 = \exp \Bigl[- {\rm Tr} \ln \Bigl( (\hat{p}_{\alpha} \gamma_{\alpha} - m) \frac{1}{\hat{p}_{\alpha} \gamma_{\alpha} - q A_{\alpha} \gamma_{\alpha} - m + i \epsilon} \Bigr) \Bigr], \nonumber\\
\end{eqnarray}
where $\epsilon$ is an arbitrary small number for the purpose of convergence. The vacuum persistence is
\begin{eqnarray}
\vert S_0 (A) \vert^2 = e^{-2 W} =  e^{-2 {\rm Im} \int d^4 x {\cal L}_{\rm eff}},
\end{eqnarray}
where
\begin{eqnarray}
W = \frac{1}{2} {\rm Tr} \int_{0}^{\infty} \frac{ds}{s} \langle x \vert e^{is [({\bf p}- q {\bf A})^2 + \frac{q}{2} \sigma_{\alpha \beta} F^{\alpha \beta}]} - e^{i s p^2} \vert x \rangle.
\end{eqnarray}

The Heisenberg-Euler and Schwinger effective action per unit four volume is given by \cite{heisenberg-euler36,schwinger51}
\begin{eqnarray}
{\cal L}_{\rm eff} &=&  - \frac{1}{2 (2 \pi)^2} \int_{0}^{\infty} \frac{ds}{s^3} e^{- m^2 s}\Bigl[ (qs)^2 {\cal G} \frac{{\rm Re} \cosh (qXs)}{{\rm Re} \sinh (qXs)} \nonumber\\ &&- 1 - \frac{2}{3} (qs)^2 {\cal F} \Bigr], \label{hes act}
\end{eqnarray}
where ${\cal F}$ is the Maxwell scalar and ${\cal G}$ is the pseudo-scalar:
\begin{eqnarray}
{\cal F} &=& \frac{1}{4} F_{\alpha \beta} F^{\alpha \beta} = \frac{1}{2} \bigl( {\bf B}^2 - {\bf E}^2\bigr), \nonumber\\
{\cal G} &=& \frac{1}{4} F^*_{\alpha \beta} F^{\alpha \beta} = {\bf B} \cdot {\bf E},
\end{eqnarray}
and $X = \sqrt{2{\cal F} + 2i {\cal G}}$.
The subtraction in Eq. (\ref{hes act}) regulates away those terms for renormalization of the vacuum energy and charge. In a pure electric field $({\bf B} = 0)$, the Schwinger effect of pair production from the imaginary part of  the QED action (\ref{hes act}) would be dominated by massless particle-antiparticle pairs if they exist since the Compton wavelength for a massless particle is infinite and any electric field can produce pair of massless charged particles and antiparticles.

The pair-production rate can be found from the Bogoliubov transformation between the in- and the-out vacua. The in-out formalism combined with the gamma-function regularization can recover the QED action (\ref{hes act}), as will be shown in the next section.
It can also be applied even to the massless QED by expressing the gamma function in terms of the zeta function \cite{kim-lee14}, which will not be reviewed in this paper.

\section{Bogoliubov Transformation}\label{sec3}

The in-state of a charge under the interaction with an external electromagnetic field and/or a curves spacetime evolves  into the out-state. These states are related to each other through the Bogoliubov transformation
\begin{eqnarray}
\hat{a}_{ {\rm out} {\kappa}} = \mu_{\kappa} \hat{a}_{ {\rm in} {\kappa}} + \nu^*_{\kappa} \hat{b}^{\dagger}_{{\rm in} {\kappa}} = \hat{U}_{\kappa} \hat{a}_{ {\rm in} {\kappa}} \hat{U}_{\kappa}^{\dagger}, \nonumber\\
\hat{b}_{ {\rm out} {\kappa}} = \mu_{\kappa} \hat{b}_{ {\rm in} {\kappa}} + \nu^*_{\kappa} \hat{a}^{\dagger}_{ {\rm out} {\kappa}} = \hat{U}_{\kappa} \hat{b}_{ {\rm in} {\kappa}} \hat{U}_{\kappa}^{\dagger},
\end{eqnarray}
where $\hat{a}_{\kappa}$ and $\hat{b}_{\kappa}$ are the annihilation operators for a particle and antiparticle carrying quantum number $\kappa$ for the field, such as energy-momentum and spin etc. The quantization rule based on the CTP theorem leads to the bosonic commutation relation
\begin{eqnarray}
[\hat{a}_{{\rm out} {\kappa}}, \hat{a}^{\dagger}_{{\rm out} {\kappa'} }] = \delta_{\kappa \kappa'}, \quad [\hat{b}_{{\rm out} {\kappa}}, \hat{b}^{\dagger}_{{\rm out} {\kappa'}}] = \delta_{\kappa \kappa'},
\end{eqnarray}
and the fermionic commutation relation
\begin{eqnarray}
\{ \hat{a}_{{\rm out} {\kappa} }, \hat{a}^{\dagger}_{{\rm out} {\kappa'} } \} = \delta_{\kappa \kappa'}, \quad \{\hat{b}_{{\rm out} {\kappa}}, \hat{b}^{\dagger}_{{\rm out} {\kappa'} } \} = \delta_{\kappa \kappa'}.
\end{eqnarray}
The mean number of produced pairs is given by
\begin{eqnarray}
N_{\kappa} = \vert \nu_{\kappa} \vert^2.
\end{eqnarray}
The Bogoliubov coefficients satisfy the relation
\begin{eqnarray}
\vert \mu_{\kappa} \vert^2 - (-1)^{2 |\sigma|} \vert \nu_{\kappa} \vert^2 =1. \label{bog rel}
\end{eqnarray}
The relation (\ref{bog rel}) lies at the root of the vacuum persistence in the next section.

For a bosonic field, the out-vacuum is the multi-particle states but unitary inequivalent representation of the in-vacuum \cite{kim-lee-yoon08}
\begin{eqnarray}
\vert 0; {\rm out} \rangle &=& \prod_{\kappa} \hat{U}_{\kappa} \vert 0; {\rm in} \rangle \nonumber\\
&=& \prod_{\kappa} \frac{1}{\mu_{\kappa}} \sum_{n_{\kappa}}
\Bigl(- \frac{\nu_{\kappa}^*}{\mu_{\kappa}} \Bigr)^{n_{\kappa}} \vert n_{\kappa}, \bar{n}_{\kappa}; {\rm in} \rangle.
\end{eqnarray}
For a fermionic field, the out-vacuum is given by
\begin{eqnarray}
\vert 0; {\rm out} \rangle &=& \prod_{\kappa} \hat{U}_{\kappa} \vert 0; {\rm in} \rangle \nonumber\\ &=& \prod_{\kappa} \Bigl( - \nu^*_{\kappa} \vert 1_{\kappa}, \bar{1}_{\kappa}; {\rm in} \rangle + \mu_{\kappa} \vert 0_{\kappa}, \bar{0}_{\kappa}; {\rm in} \rangle \Bigr).
\end{eqnarray}
The Pauli blocking suppresses production of more than one pair in the same state.

\section{In-Out Formalism for Effective Action} \label{sec4}

The Schwinger variational principle leads to the effective action for a particle in an electromagnetic field and/or curved spacetime  \cite{schwinger51b,schwinger51c}
\begin{eqnarray}
e^{iW} = \langle 0, {\rm out} \vert  0, {\rm in} \rangle, \label{eff-def}
\end{eqnarray}
where $W$ is the integrated action
\begin{eqnarray}
W = \int \sqrt{-g} d^4 x {\cal L}_{\rm eff}. \label{int act}
\end{eqnarray}
The energy-momentum tensor is the variation of $W$ with respect to the spacetime metric \cite{birrell-davies,dewitt03}.
In the in-out formalism, the integrated action can be expressed in terms of the scattering matrix as
\begin{eqnarray}
W = - i \ln (\langle 0, {\rm out} \vert 0, {\rm in}
\rangle) = (-1)^{2 |\sigma|} i {\cal V} \sum_{\kappa} \ln (\mu_{\kappa}^*), \label{bog-eff}
\end{eqnarray}
where a factor ${\cal V}$ is introduced to make the summation dimensionless.
The one-loop effective action is equivalent to the sum of all Feynman diagrams in Fig. 2.
In fact, the summation of all Feynman diagrams in a constant electric field was attempted in Ref. \cite{chiu-nussinov}, but it is not practical to proceed in this direction. The one-loop effective action may become complex due to particle production from the background field. The most advantageous point of the in-out formalism is the consistent relation between the complex effective action and the vacuum persistence for particle production
\begin{eqnarray}
&&\vert \langle 0, {\rm out} \vert 0, {\rm in} \rangle \vert^2 = e^{- 2{\rm Im} W}, \nonumber\\
&&2 {\rm Im} W = (-1)^{2 |\sigma|} {\cal V} \sum_{\kappa} \ln( 1 -(-1)^{2 |\sigma|} N_{\kappa}). \label{vac per}
\end{eqnarray}
\begin{figure} \label{fig2}
\includegraphics[width=2.5in]{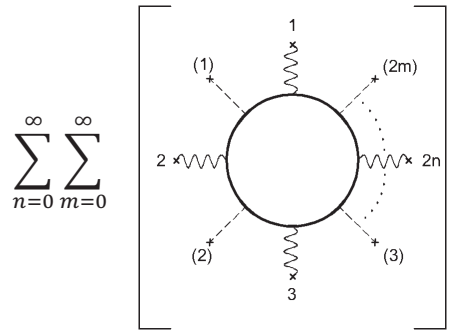}
\caption{The Feynman diagrams for the one-loop effective action. The wavy lines denote the virtual photons and the dashed lines denote the virtual gravitons.}
\end{figure}

Kim, Lee and Yoon have introduced the so-called gamma-function regularization method \cite{kim-lee-yoon08,kim-lee-yoon10a,kim11}
\begin{eqnarray}
{\rm W} = \pm i \sum_{\kappa} \ln \mu_{\kappa}^* = \pm i \sum_{\lambda} \sum_{\kappa} \ln \Gamma(a_{\lambda} + i b_{\lambda} (\kappa)). \label{gamma}
\end{eqnarray}
In the known cases $a_{\lambda} = 0, \pm 1/2, \pm 1$. The gamma-function regularization originates in part from the proper-time integral
\begin{eqnarray}
\ln \Gamma(a + i b) = \int_{0}^{\infty} \frac{ds}{s} e^{- i bs} \Bigl( \frac{e^{-as}}{1- e^{-s}}- \cdots \Bigr). \label{prop int}
\end{eqnarray}
The final expression of the effective action should involve the renormalized quantities, which can be done by the Schwinger substraction scheme \cite{schwinger51}, which discards any divergent terms in Eq. (\ref{gamma}). The contour integral or the wick-rotation in the complex plane results in the vacuum persistence (imaginary part) as well as the vacuum polarization (real part) \cite{kim12}. In the next section we shall apply the gamma-function regularization to find the QED action in a constant electric with or without a parallel magnetic field \cite{kim-lee-yoon08,kim-lee-yoon10a,kim11}.

\section{$\Gamma$-Function Regularization}\label{sec6}

In this section we apply the gamma-function regularization method to compute the QED action in a constant electric field. The QED action in an electric field is interesting since the vacuum becomes unstable against the pair production and the quantum field theory should be able to properly handle the Schwinger effect. The in-out formalism has proven a consistent field theoretical method in this sense.

Let us consider the Dirac equation for a spin-1/2 fermion in an electric field $E(t)$ with a fixed direction. It has the momentum- and spin-state component equation
\begin{eqnarray}
\bigl[ \partial_t^2 + m^2+ {\bf p}_{\perp}^2 + (p_{\parallel} + qA_{\parallel})^2+ 2 i \sigma E(t) \bigr] \phi_{\sigma {\bf p}_{\perp}} = 0.
\end{eqnarray}
In the case of a constant electric field, the vector potential is $A_{\parallel} = -E t$. The asymptotic positive frequency solutions at $t \rightarrow = \pm \infty$ are given by the parabolic function \cite{kim-lee-yoon08}
\begin{eqnarray}
 \phi^{(-\infty)}_{\sigma {\bf p}_{\perp}} &=& D_{\kappa} (\zeta), \nonumber\\
 \phi^{(\infty)}_{\sigma {\bf p}_{\perp}} &=& e^{-i \pi \kappa} D_{\kappa} (-\zeta) + \frac{\sqrt{2 \pi}}{\Gamma (-\kappa)} e^{-i \frac{\pi}{2} (\kappa+1)} D_{-\kappa-1} (i \zeta), \nonumber\\
\end{eqnarray}
where
\begin{eqnarray}
\zeta = \sqrt{\frac{2}{qE}} e^{i \frac{\pi}{4}} (p_{\parallel} - qEt), \quad \kappa = \frac{2 \sigma -1}{2} - i \frac{m^2 + {\bf p}_{\perp}^2 }{2qE}.
\end{eqnarray}
Then, the Bogoliubov coefficients are
\begin{eqnarray}
\mu_{{\bf p}_{\perp}} = \frac{\sqrt{2 \pi e^{- i \pi (\kappa+1)}}}{\Gamma (- \kappa)}, \quad \nu_{{\bf p}_{\perp}}= e^{- i \pi \kappa }.
\end{eqnarray}
The mean number of produced pairs is
\begin{eqnarray}
N_{{\bf p}_{\perp}} = \vert \mu_{{\bf p}_{\perp}} \vert^2 = e^{- \pi \frac{m^2+  {\bf p}_{\perp}^2 }{qE}}.
\end{eqnarray}

After summing over spin states, the logarithm of the gamma function, (\ref{prop int}), takes the form
\begin{eqnarray}
\ln \Gamma ( - \kappa^*) = \int_{0}^{\infty} \frac{ds}{s} \coth \Bigl(\frac{s}{2} \Bigr) e^{i \frac{m^2 + {\bf p}_{\perp}^2}{2qE} s}. \label{con int}
\end{eqnarray}
Applying the Cauchy theorem along the contour in the Fig. 3, one obtains the following result
\begin{eqnarray}
\ln \Gamma ( - \kappa^*) &=& -i P \int_{0}^{\infty} \frac{ds}{s} \cot \Bigl(\frac{s}{2} \Bigr) e^{- \frac{m^2+ {\bf p}_{\perp}^2}{2qE} s}
\nonumber\\
&&+ \sum_{n = 1}^{\infty} \frac{1}{n} e^{- \pi \frac{m^2+ {\bf p}_{\perp}^2}{qE}n}. \label{com act}
\end{eqnarray}
Finally, the effective action (\ref{bog-eff}), after taking into account the density of states, gives
the complex action
\begin{eqnarray}
{\cal L}_{\rm eff} &=& \frac{1}{2} \Bigl( \frac{qE}{2 \pi}\Bigr)^{2} P \int_{0}^{\infty} \frac{ds}{s^2} e^{- \frac{m^2}{qE} s} \Bigl(\cot s - f_{\frac{1}{2}} (s) \Bigr)\nonumber\\
&& - i \Bigl( \frac{qE}{2 \pi} \Bigr) \int \frac{d^{2}
{\bf p}_{\perp}}{(2 \pi)^{2}} \ln \Bigl( 1 - e^{- \pi \frac{m^2+ {\bf p}_{\perp}^2}{qE}} \Bigr). \label{com act2}
\end{eqnarray}
Here, $P$ denotes the principal value and $f_{\frac{1}{2}} = 1/s - s/3$ subtracts two divergent terms in the principal integral and renormalizes the vacuum energy and charge.
\begin{figure}
\includegraphics[width=3.0in]{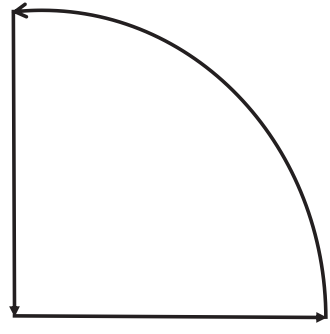} \label{fig3}
\caption{The contour for Eq. (\ref{con int}) in the first quadrant, which gives the principal value and the imaginary part for the vacuum persistence in Eq. (\ref{com act2}).}
\end{figure}

We now consider spin-1/2 fermion in a constant electric and a constant parallel magnetic field. The Dirac equation projected into spin-eigenstates gives the Bogoliubov coefficients \cite{kim12b}
\begin{eqnarray}
\mu^{(\sigma)}_{(r)n} = \frac{\sqrt{2 \pi e^{- i \pi (2 \kappa^*+ \kappa+1)}}}{\Gamma (- \kappa)}, \quad \nu^{(\sigma)}_{(r)n} = e^{- i \pi \kappa^*},
\end{eqnarray}
where $n$ is the Landau level, $\sigma = \pm 1/2$ is the spin eigenvalue along the magnetic field, $r = \pm 1$ is another spin eigenvalue along the electric field, and
\begin{eqnarray}
\kappa = - \frac{r+1}{2} + i \frac{m^2 + qB (2n+1 - 2 \sigma)}{2qE}.
\end{eqnarray}
The effective action and the Schwinger effect follow from the spin-averaged Bogoliubov coefficients
\begin{eqnarray}
\mu^{(\sigma)}_{n} = \bigl(\mu^{(\sigma)}_{(1)n} \mu^{(\sigma)}_{(-1)n}\bigr)^{\frac{1}{2}}, \quad \nu^{(\sigma)}_{n} = \bigl(\nu^{(\sigma)}_{(1)n} \nu^{(\sigma)}_{(-1)n}\bigr)^{\frac{1}{2}}.
\end{eqnarray}
Then, the mean number of produced pairs in the $n$-th Landau level and the spin $\sigma$ is
\begin{eqnarray}
N^{(\sigma)}_{n} = \vert \nu^{(\sigma)}_{n} \vert^2 = e^{- \pi \frac{m^2 + qB (2n+1 - 2 \sigma)}{qE}}.
\end{eqnarray}

The effective action follows from Eq. (\ref{bog-eff}) as
\begin{eqnarray}
{\cal L}_{\rm eff} = - i \Bigl( \frac{qE}{2 \pi} \Bigr) \Bigl( \frac{qB}{2 \pi} \Bigr) \sum_{\sigma n} \ln \bigl(\mu^{(\sigma)*}_{n} \bigr).
\end{eqnarray}
After summing over the spin and Landau levels, one obtains the unrenormalized effective action
\begin{eqnarray}
{\cal L}_{\rm eff} = \frac{1}{2} \Bigl( \frac{qE}{2 \pi} \Bigr) \Bigl( \frac{qB}{2 \pi} \Bigr)  \int_{0}^{\infty} \frac{ds}{s}\coth \Bigl(\frac{s}{2} \Bigr) \cot \Bigl(\frac{Bs}{2E} \Bigr) e^{- i \frac{m^2}{2qE} s}. \nonumber\\
\end{eqnarray}
Finally, the contour integral in the fourth quadrant gives the complex effective action \cite{kim12b}
\begin{eqnarray}
{\cal L}_{\rm eff} &=& - \frac{1}{2 (2 \pi)^2} P \int_{0}^{\infty} \frac{ds}{s^3}e^{- m^2 s} \nonumber\\ && \times  \Bigl[ \frac{(qBs) (qEs)}{\tanh (qBs) \tan (qEs)} - 1 - \frac{(qs)^2}{3} (B^2 - E^2)\Bigr] \nonumber\\
&&+ \frac{i}{2} \Bigl( \frac{qE}{2 \pi} \Bigr) \Bigl( \frac{qB}{2 \pi} \Bigr) \sum_{k = 1}^{\infty} \frac{1}{k} e^{-\frac{\pi m^2}{qE}k} \coth \Bigl(\frac{\pi B}{E} k \Bigr). \label{EB com act}
\end{eqnarray}
One can show that the vacuum persistence still holds
\begin{eqnarray}
2 {\rm Im} {\cal L}_{\rm eff} = - \Bigl( \frac{qE}{2 \pi} \Bigr) \Bigl( \frac{qB}{2 \pi} \Bigr) \sum_{\sigma n} \ln \bigl( 1-  N^{(\sigma)}_{n} \bigr). \label{EB vac per}
\end{eqnarray}

\section{Reconstruction of Effective Action} \label{sec5}

One can obtain the pair-production rate from the effective action provided that the action has an imaginary part due to the instability of the Dirac vacuum. The QED action by Heisenberg, Euer and Schwinger, for instance, has poles in the proper-time integral. It is a consequence of the Cauchy theorem. In fact, the mean number of produced pairs from the vacuum is related to the vacuum persistence as shown in Eq. (\ref{vac per}). Now, one may raise another question whether one can find the effective action from the pair-production rate. The optical dispersion relation in particle physics is the most well-known example. There are various methods to compute the pair-production rate or the mean number under the influence of a strong electric field in QED, chromelectric field in QCD, black holes and expanding universe etc. Therefore, the inverse procedure will provide strong field physics with a field theoretical method useful for understanding the vacuum polarization in strong field, which belongs to nonperturbative quantum regime. So, any systematic method in this direction will open a new window to the regime of nonperturbative quantum phenomena.

Kim and Schubert seriously considered the reconstruction of the effective action from the pair production \cite{kim-taida}.
If the argument of the logarithmic function of the imaginary part of the effective action can be factorized into a product of one plus or one minus exponential factors, then the structure of simple poles and their residues of these factors uniquely determine the analytical structure of the proper-time integrand of the effective action:
\begin{eqnarray}
2 {\rm Im} {\cal L}_{\rm eff} = (-1)^{2 |\sigma|} \sum_{\kappa} \ln \bigl( 1 + (-1)^{2 |\sigma|} N_{\kappa} \bigr).
\end{eqnarray}
The distribution function of produced particles in known physics has the general form
\begin{eqnarray}
N_{\kappa} = \frac{1}{e^{\beta \omega_{\kappa}} + \cos\theta}
\end{eqnarray}
where $\cos \theta = 1$ for the Fermi-Dirac (FD) distribution, $\cos \theta = -1$ for the Bose-Einstein (BE) distribution and $\cos \theta = 0$ for the Boltzmann (B) distribution. Then, the vacuum persistence becomes \cite{kim16a,kim16b}
\begin{eqnarray}
2 {\rm Im} {\cal L}_{\rm FD/BE} = (-1)^{2 |\sigma|+1} \sum_{\kappa} \ln \bigl( 1 + (-1)^{2 |\sigma|+1} e^{- \beta \omega_{\kappa}}  \bigr), \nonumber\\
\end{eqnarray}
and
\begin{eqnarray}
2 {\rm Im} {\cal L}_{\rm B} = (-1)^{2 |\sigma|} \sum_{\kappa} \ln \bigl( 1 + (-1)^{2 |\sigma|} e^{- \beta \omega_{\kappa}}  \bigr). \label{B vac per}
\end{eqnarray}

The conjecture by Kim and Schubert is that the effective action may be reconstructed by the vacuum persistence as \cite{kim-taida,kim16a,kim16b}
\begin{eqnarray}
{\cal L}_{\rm FD/BE} &=& (-1)^{2 \sigma + 1} \sum_{\kappa}  P \int_{0}^{\infty} \frac{ds}{s} e^{-\frac{\beta \omega_{\kappa}}{\pi}s} \nonumber\\&& \times \Bigl[ \frac{(\cos s)^{1-2 |\sigma|}}{\sin s} - f_{\frac{1}{2} - |\sigma|} (s) \Big], \label{FD-BE}
\end{eqnarray}
and
\begin{eqnarray}
{\cal L}_{\rm B} = (-1)^{2 |\sigma| } \sum_{\kappa}  P \int_{0}^{\infty} \frac{ds}{s} e^{-\frac{\beta \omega_{\kappa}}{\pi}s} \Bigl[ \frac{(\cos s)^{2 |\sigma|}}{\sin s} - f_{|\sigma|} (s) \Bigr], \nonumber\\ \label{B}
\end{eqnarray}
where the function $f_{|\sigma|}$ corresponds to the Schwinger subtraction scheme to make the action finite. In QED the subtracted terms correspond to the renormalization of the vacuum energy and charge.

The argument leading to Eqs. (\ref{FD-BE}) and (\ref{B}) is the Mittag-Leffler's theorem \cite{markushevich}. Simply stated, the theorem reads that let $b_n, (n=1, 2, \cdots)$ be a sequence of complex numbers satisfying $|b_n| \rightarrow \infty$ as $n \rightarrow \infty$ and let $g$  and $h$  be two meromophic functions having simple poles at $b_1, b_2, \cdots$ such that ${\rm Res} [g, b_n] = {\rm Res} [h, b_n]$ for $n = 1, 2, \cdots$. Then, $g-h$ is an entire function.
As the Mittag functions (entire functions, such as  $e^{z^2}$) are independent of renormalization, it is highly likely that they are removed through renormalization and only those functions with simple poles are physically relevant ones for the vacuum polarization \cite{kim-taida}.

\section{QED Actions from Reconstruction Conjecture} \label{sec7}

We reconstruct the effective action from the conjecture by Kim and Schubert in section \ref{sec5}. As will be shown below, the reconstruction of the
effective action is simple and useful provided that the pair production is known by some means.

First, in the case of a constant electric field, the complex action gives the vacuum persistence
\begin{eqnarray}
2 {\rm Im} {\cal L}_{\rm eff} = -2 \Bigl( \frac{qE}{2 \pi} \Bigr) \int \frac{d^{2}
{\bf p}_{\perp}}{(2 \pi)^{2}} \ln \Bigl( 1 - e^{- \pi \frac{m^2+ {\bf p}_{\perp}^2}{qE}} \Bigr).
\end{eqnarray}
Then, Eqs. (\ref{B vac per}) and (\ref{B}) imply the effective action
\begin{eqnarray}
{\cal L}_{\rm eff} &=& \Bigl( \frac{qE}{2\pi} \Bigr) \int \frac{d^{2}
{\bf p}_{\perp}}{(2 \pi)^{2}} P \int_{0}^{\infty} \frac{ds}{s} e^{- \frac{m^2 + {\bf p}_{\perp}^2 }{qE}s} \nonumber\\
&&\times \Bigl[\cot(s) - \frac{1}{s} + \frac{s}{3} \Bigr]. \nonumber\\
\end{eqnarray}
Second, in the case of a parallel electric and magnetic field, the vacuum persistence (\ref{EB vac per}) implies the effective action
\begin{eqnarray}
{\cal L}_{\rm eff} &=& - \frac{1}{2} \Bigl( \frac{qE}{2 \pi} \Bigr) \Bigl( \frac{qB}{2 \pi} \Bigr) \sum_{\sigma n} \int_{0}^{\infty} \frac{ds}{s} e^{- \frac{m^2 + qB (2n+1 - 2 \sigma)}{qE}s } \nonumber\\ && \times \Bigl[ \cot s - \frac{1}{s} + \frac{s}{3} \Bigr] \nonumber\\
&=& - \frac{1}{2} \Bigl( \frac{qE}{2 \pi} \Bigr) \Bigl( \frac{qB}{2 \pi} \Bigr) \int_{0}^{\infty} \frac{ds}{s} e^{- \frac{m^2}{qE}s } \nonumber\\
&& \times \Bigl[ \coth \Bigl(\frac{B}{E}s \Bigr) \cot s - \frac{E}{B s^2} + \frac{B^2 - E^2}{3}  \Bigr].
\end{eqnarray}
Here, the dots denote the Schwinger substraction scheme for renormalization of the vacuum energy and charge. As illustrated for the Schwinger effect in a constant electric field with or without a constant parallel magnetic field, the information of the mean number straightforwardly leads to the one-loop effective action.

\section{Conclusion}

Physics in a strong field exhibits nonperturbative effects, such as the Schwinger effect of pair production in an electric field and the Hawking radiation of particles in a black hole. Spontaneous particle production requires a consistent framework for quantum field theory since the vacuum becomes unstable against particle production. To explore the nonperturbative effect, one should be able to calculate the one-loop effective action, whose real part is the vacuum polarization, and whose imaginary part is the vacuum persistence. We have shown that the in-out formalism provides complex one-loop effective actions, satisfying the consistent requirement. In particular, the in-out formalism combined with the gamma-function explicitly gives the exact effective action once the Bogoliubov coefficient is known in terms of the gamma functions. We have also advanced a conjecture for reconstruction of the effective action from the knowledge of the pair production. We have illustrated the QED action in a constant electric and magnetic field. The reconstruction conjecture will provide a very useful field theoretical method in various areas of physics.

\section*{Acknowledgements}
This work was supported in part by Basic Science Research Program through the National Research Foundation of Korea(NRF) funded by the Ministry of Education(2015R1D1A1A01060626) and by Institute for Basic Science (IBS) under IBS-R012-D1.

\end{document}